\documentstyle[aps,epsf,prl]{revtex}
\begin{document}
\twocolumn[\hsize\textwidth\columnwidth\hsize\csname
@twocolumnfalse\endcsname

\draft
\title{On Metric Preheating}
\author{K. Jedamzik\cite{byline1}}
\address{Max-Planck-Institut f\"ur Astrophysik,
Karl-Schwarzschild-Str. 1, 85740 Garching, Germany} 
\author{G. Sigl\cite{byline2}}

\address{University of Chicago, Department of Astronomy and
Astrophysics, 5640 South Ellis Avenue, Chicago, IL 60637\\
DARC, UMR--8629, CNRS, Observatoire de Paris-Meudon, F-92195
Meudon C\'{e}dex, France}

\maketitle

\begin{abstract}
We consider the generation of super-horizon metric fluctuations
during an epoch of preheating in the presence of a scalar field $\chi$
quadratically coupled to the inflaton. We find that the requirement
of efficient broad resonance is concomitant with a severe damping
of super-horizon $\delta\chi$ quantum fluctuations during
inflation.  Employing perturbation theory with backreaction included
as spatial averages to second order in the scalar fields and in the metric,
we argue that 
the usual inflationary prediction for metric perturbations on scales relevant
for structure formation is not strongly modified.
\end{abstract}

\pacs{PACS numbers: 98.80.Cq, 04.25.Nx}
\vskip2.2pc]

{\it Introduction.}
In the inflationary paradigm the universe went through an early and extended
period of accelerated expansion $\ddot{a}>0$ with rapid growth
of the scale factor $a$. Inflation may provide solutions to a variety of
cosmological \lq\lq problems\rq\rq\, as well as making generic predictions
for a scale-invariant Harrison-Zel'dovich spectrum of density fluctuations,
possibly acting as seeds for structure formation~\cite{reviews,MFB92}. After an 
inflationary period cosmic energy density is dominated by oscillations of 
the inflaton $\varphi$. Efficient conversion of this energy to other coupled 
bosons by parametric resonance phenomena may occur
during a period of preheating immediately following the
end of inflation~\cite{PRE,KLS97}.
Consider, for example, a chaotic inflationary scenario with Lagrange density
${\it L} = {1\over 2}\varphi_{;\mu} \varphi^{;\mu} - V(\varphi ,\chi )$
and potential
\begin{equation}
\label{potential}
V(\varphi ,\chi ) = {m^2\over 2}\varphi^2 + {g^2\over 2}\varphi^2\chi^2\, , 
\end{equation}
where $\chi$ is a boson coupled to the inflaton $\varphi$. After the
end of slow roll inflation, i.e. for $\varphi\lesssim (12\pi)^{-1/2}$,
the linearized equation of motion for the comoving Fourier mode
$k=|{\bf k}|$ of the $\chi$ field is,
\begin{equation}
\label{Mathieu}
\delta\ddot{\chi}_k + 3H\delta\dot{\chi}_k +\left[\left({k\over a}\right)^2
+ g^2{\Phi}^2{\rm sin}^2(mt)\right]\delta\chi_k = 0\, ,
\end{equation}
where the amplitude $\Phi$ of the sinusoidal inflaton oscillations,
$\varphi = \Phi\, {\rm sin}(mt)$, decays as
$\Phi =\Phi (t_0)(a/a_0)^{-3/2}$, due to the Hubble expansion of the
universe which is governed by the Friedmann equation
\begin{equation}
  H^2\equiv\left(\frac{\dot a}{a}\right)^2=\frac{\kappa^2}{3}
  \left[\frac{1}{2}\dot\varphi^2+\frac{1}{2}\dot\chi^2+
  V(\varphi,\chi)\right]\label{Friedmann}
\end{equation}
for the scale factor $a(t)$, 
where $\kappa^2=8\pi /M_{pl}^2$, with $M_{pl}$ the Planck mass.

Eq. (\ref{Mathieu}) is essentially known as the Mathieu equation which 
exhibits parametric resonance,
i.e. exponentially growing modes for $\delta\chi_k$, due to the time-dependence
of the $\chi$ mass. One may associate a resonance-parameter
$q(t)=g^2\Phi(t)^2/4m^2$ with the system. In an expanding universe,
parametric resonance is efficient for $q>1$ where resonance occurs
for wavevectors $k$ within
broad bands, and is rendered inefficient for $q\ll 1$~\cite{KLS97}.
In the broad resonance regime $q>1$, exponential growth of bosonic
fluctuations is possible for super-horizon modes even in the limit
$k\to 0$. Recently, the important realization has been made that 
growth of coherent, bosonic matter fluctuations during preheating
is accompanied by production of metric fluctuations possibly leading
to abundant primordial black hole formation~\cite{BTKM99}. It has also
been pointed out that a significant modification of the power spectrum of 
density fluctuations relevant for structure formation 
and cosmic microwave background anisotropies is
conceivable~\cite{BTKM99,FB99}.

{\it Multiple scalar fields and metric perturbations.}
Consider perturbations around a Robertson-Walker metric in
longitudinal gauge
\begin{equation}
\label{metric}
ds^2 = (1 + 2\phi)dt^2 - a^2(t)(1-2\phi)d{\bf x}^2\, ,
\end{equation}
where $\phi$ is a gauge-invariant potential, quantifying the
density perturbation $\delta\rho /\rho$
at horizon-crossing in unperturbed Hubble flow. 
The Fourier-transformed, first-order Einstein equations 
for coupled bosons $\varphi_I(t)$ and their perturbations 
$\delta\varphi_{I}(t,{\bf x})$ give~\cite{KP88}
\begin{eqnarray}
\label{extra}
3H\dot{\phi}_k + \left[(k/a)^2 + 3H^2\right]{\phi}_k = \nonumber \\
-{\kappa^2\over 2}\sum_I\bigl(\dot{\varphi}_I \delta\dot{\varphi}_{Ik}
-{\phi}_k\dot{\varphi}^2_I + V_{\varphi_I}\delta\varphi_{Ik}\bigr) \, ,\\ 
\label{delphi}
\delta\ddot{\varphi}_{Ik} + 3H\delta\dot{\varphi}_{Ik} +\left[(k/a)^2 
+V_{\varphi_I\varphi_I}\right]\delta\varphi_{Ik} = \nonumber \\
4\dot{\phi}_k\dot{\varphi}_I +
2\bigl(\ddot{\varphi}_I+3H\dot{\varphi}_I\bigr)\phi_k - \sum_{J\not= I}
V_{\varphi_J\varphi_I}\delta\varphi_{Jk}\, , \\
\label{constraint}
\dot{\phi}_k + H\phi_k =
{\kappa^2\over 2}\sum_I \dot\varphi_I\delta\varphi_{Ik} \, ,
\end{eqnarray}
where a subscript $\varphi_I$ denotes derivative with respect to $\varphi_I$.
Eq. (\ref{extra}) and Eq. (\ref{constraint})
may be combined to
\begin{equation}
\label{explicit}
\phi_k = -\frac{\kappa^2}{2}
\frac{\sum_I\bigl(\dot{\varphi}_I\delta\dot{\varphi}_{Ik}
+3H\dot{\varphi}_I\delta\varphi_{Ik}
+V_{\varphi_I}\delta\varphi_{Ik}\bigr)}
{(k/a)^2-(\kappa^2/2)\sum_I\dot{\varphi}_I^2}\, ,
\end{equation}
explicitly showing that $\phi$ is completely determined when the evolution of
the matter fields is known.

Whether or not parametric resonance for $k\to 0$ induces significant
secondary perturbations in $\phi$ is largely a question of initial
conditions. The evolution equations for the unperturbed, background fields
are $\ddot{\varphi}_I + 3H\dot{\varphi}_I + V_{\varphi_I}=0$.
Consider a chaotic inflationary period with inflaton potential as in
Eq.~(\ref{potential}). Assuming that inflation is driven by $\varphi$
(i.e. $g^2\chi^2\ll m^2$) we have
$\dot{\varphi} = -(m^2/3H)\varphi$ by the slow-roll approximation
for $\varphi >\varphi_0\simeq(12\pi)^{-1/2}$.
In the limit $g\varphi \gg H$
the evolution of the homogeneous $\chi$ field may be derived in the
adiabatic approximation $\chi\propto|g\varphi|^{-1/2}a^{-3/2}{\rm cos}\int
g\varphi dt$, such that the
amplitude of $\chi$ is decaying during a phase of exponential growth of
the scalefactor. The adiabatic approximation holds 
if $q_{0}^{-1/2}(\varphi /\varphi_0)^{-2} \ll 2$, quite generally when the
resonance parameter at the beginning of preheating
satisfies $q_0\gtrsim1$. The amplitudes of fluctuations in the
$\chi$ field similarly decay as $\delta\chi_k\propto
\omega_{\chi_k}^{-1/2}a^{-3/2}\cos\int\omega_{\chi_k} dt$
where $\omega_{\chi_k}^2=
(k/a)^2+(g\varphi)^2$ is the oscillation frequency of mode $k$. This
can be seen by noting that the R.H.S. of
Eq.~(\ref{delphi}) contains factors of $\chi$ and $\dot\chi$
whose values are strongly suppressed compared to $\varphi$
and $\dot\varphi$ due to the inflationary expansion.
When interpreted classically, the decay of the amplitude of the 
$\delta\chi_k$ fluctuations corresponds to a dilution of particle number
densities associated with the $\delta\chi_k$ field by the expansion of the
universe. In contrast, 
super-horizon fluctuations in the inflaton field $\delta\varphi_k$ are not
damped, i.e. $\delta\varphi_k\simeq constant$, 
and lead to the usual prediction of a scale-invariant
Harrison-Zel'dovich spectrum of density fluctuations. 
Hawking radiation in the de Sitter phase of an inflationary universe
enhances the $\delta\varphi_k$ amplitude for superhorizon modes, which
is possible since the mass of the inflaton $m$ is somewhat smaller than
the Hubble constant $H$. Demanding 
$M_{\chi}=g\varphi\lesssim H$, in order to have a similar effect for 
$\delta\chi_k$ fluctuations, we arrive at the condition
$q_{0}\lesssim1/6^{1/2}$. In particular, damping seems inevitable
for systems in the broad resonance regime.

Consider the amplitudes of vacuum quantum fluctuations in the bosonic
fields on sub horizon scales $k/a\gtrsim H$. These may be estimated
from quantum field theory in flat space where
the $\delta\varphi_I$'s are replaced by operators
$\delta\widehat{\varphi}_I(t,{\bf x}) = \int(d^3{\bf k})/(2\pi)^3
[\widehat{a}_{\bf k}\delta\widetilde{\varphi}_{I{\bf k}} 
e^{-i{\bf kx}} +H.C.]$, with $\widehat{a}_{\bf k}$ the usual
annihilation operators
and $H.C.$ denoting the hermitian conjugate. A typical amplitude may be
estimated by the square-root of the expectation value of
$\widehat{\varphi}_I^2(t,{\bf x})$ in the vacuum. This gives
\begin{equation}
\label{vac}
\langle 0| \delta\widehat{\varphi}_I^2(t,{\bf x}) |0 \rangle =\int
\frac{d^3{\bf k}}{(2\pi)^3}|\delta\widetilde{\varphi}_{I{\bf k}}|^2
\simeq\int\frac{d^3({\bf k}/a)}{(2\pi)^3\omega_{Ik}}\, ,
\end{equation}
where $\omega_{Ik}^2 = (k/a)^2 + M_{\varphi_I}^2$.
Therefore, the vacuum amplitudes in a comoving wavevector range $k$
are given by
\begin{equation}
  \left|\delta\varphi_{Ik0}\right|^2\equiv
  k^3\left|\delta\widetilde{\varphi}_{Ik}\right|^2
  =\frac{(k/a)^3}{2\pi^2\omega_{Ik}}\,.\label{vacampli}
\end{equation}
At horizon exit of a mode,
i.e. at $k/a = H$, perturbations in the inflaton therefore have
$\delta\varphi_k|_{ex}\simeq H/(2\pi^2)^{1/2}$ since $m\lesssim H$.
Similarly, for the massive field $\delta\chi_k$ we find
$\delta\chi_k|_{ex}\simeq H^{3/2}/|g\varphi|^{1/2}/(2\pi^2)^{1/2}$.
In what follows we will compute the evolution of the boson fields by the
classical equations which is a good approximation in the limit of large
occupation numbers, in particular, when the amplitudes are much larger than
their vacuum expectation values.

Let us evaluate the wavevectors for modes which contribute to the
formation of large-scale structure. 
Let us define, $T_0\equiv (m\varphi_0)^{1/2}$, which is the
approximate reheat temperature, $T_{RH}\simeq T_0$, if reheating would
be instantaneous. For what follows it is sufficient to approximate reheating
to be instantaneous, such that $a$ evolves as $a\simeq (t/t_0)^{1/2}$ for
$t\gg t_0$ [setting $a(t_0)\equiv1$]. Then the proper horizon
distance evolves as  $r_H^p(T)\simeq r_H^p(T_0)(T_0/T)^2$ with
$r_H^p(T_0)\simeq M_{pl}/T_0^2$, the horizon distance 
immediately after inflation. A mode with comoving
wave vector $k\simeq (T_{re}/T_0)/r_H^p(T_0)\simeq (m/3)(T_{re}/T_0)$ 
re-enters into the horizon at $T_{re}\ll T_0$.
This mode left the horizon during inflation at time $t_{ex}$ when
$k/a_{ex}\simeq H(\varphi_{ex})$. This allows one to derive the scale factor 
at horizon exit, $a_{ex}\simeq (T_{re}/T_0)(\varphi_{ex}/\varphi_0)^{-1}$.
For $k/a\ll g\varphi$ the amplitude of the matter field at the
beginning of preheating $\delta\chi_k(t_0) =\delta\chi (t_{ex})
a_{ex}^{3/2}(\varphi_{ex}/\varphi_0)^{1/2}$
is given by
\begin{eqnarray}
\label{amplitude}
\delta\chi_k(t_0) 
\simeq 0.1mq_0^{-1/4}(T_{re}/T_0)^{3/2}
(\varphi_{ex}/\varphi_0)^{-1}.
\end{eqnarray} 
For small $k$ and low $T_{re}\sim 1$eV this is very small  
compared to a typical inflaton perturbation $\delta\varphi_k\sim m$.

Non-perturbative decay of the inflaton condensate into bosonic matter
fields via parametric resonance occurs when the condition of adiabaticity
for the $\chi$ field is violated, i.e. when
$\dot{\omega}_{\chi_k}\gtrsim\omega_{\chi_k}^2$.
In what follows, we follow closely the analytic treatment by Kofman, Linde,
and Starobinsky~\cite{KLS97} (KLS97)
for the stochastic growth of bosonic particle number in the
broad resonance parameter regime, $q\gg 1$, and in an expanding universe.
The largest bosonic wavevector $k_{max}$ 
which receives parametric amplification
is given by $k_{max}/a \simeq (gm\Phi)^{1/2}/2$. Due to phase space
arguments, this is also the
typical wavevector $k_*$ in which most of the inflaton energy is \lq\lq
dumped\rq\rq . Let us define $\delta X_k$
as the amplitude of the boson fields, i.e. $\delta X_k = \delta\chi_k$ at
moments when $\delta\dot{\chi}_k = 0$. KLS97 were able to derive that
the comoving occupation number $n_k$ evolves as,
$2n_k=\omega_{\delta\chi_k}|\delta \tilde{X}_k|^2 a^3 -1\simeq 
{\rm exp}(2m\int dt\mu_k(t))$, where the integral may be replaced by
$2m\mu t$, with an effective value $\mu\simeq 0.1-0.18$. Replacing
the Floquet index by an effective index accounts for the stochasticity
of the resonance, due to wavevectors shifting in and out of resonance bands
with the expansion of proper wavelength and the time-dependency of the
resonance parameter $q(t)$. Note that there is still some dependency of 
$\mu$ on $k$, which implies that the decay of the
inflaton energy is dominated by decay into modes with $k$ in a
narrow neighborhood of the maximum unstable mode.
These considerations allowed KLS97 to estimate the number of
inflaton oscillations $N=mt/(2\pi )$ between 
$t_0\lesssim t \lesssim t_1$, where $t_1$ is the time when backreaction
effects of the produced matter particles on the inflaton become
significant, i.e. when $m^2\simeq g^2X_{rms}^2(t_1)$, where in the
following for any function $f({\bf x})$ we define $f_{rms}\equiv
\langle f^2({\bf x})\rangle^{1/2}$.
They found,
\begin{equation}
\label{N1}
N_1\simeq {1\over 8\pi \mu}\ln\left[{10^6m(2\pi N_1)^3\over
g^5M_{pl}}\right]\, .
\end{equation}
During this time the scale-factor evolves as
$a(t) = (t/t_0)^{2/3} \simeq (\pi N)^{2/3}$, such that at $t_1$ the resonance
parameter is $q_1 = g^2\Phi^2(t_1)/(4m^2)\simeq q_0/(\pi N_1)^2$.
Subsequently, if $q_1>1/4$, the inflaton oscillation frequency increases to
$m^2_{\varphi,{\rm eff}}\simeq g^2 X_{rms}^2$,
and Hubble expansion effects are negligible, i.e. $m_{\varphi,{\rm eff}}\gg H$.
The second stage of preheating, $t > t_1$, is well approximated by parametric
resonance according to the Mathieu equation with resonance parameter
$q_{\rm eff}\simeq g^2\Phi^2/(4g^2 X_{rms}^2)$. Broad resonance is 
terminated at some time $t_2$, when $q_{\rm eff}$ decreased to $\simeq 1/4$.
At this time $\Phi^2(t_2)\simeq X_{rms}^2(t_2)$. During the second stage
of preheating, the inflaton typically makes only a few oscillations,
$N_2\simeq (1/4\pi\mu)\ln4q_1^{1/4}$. Combining adiabatic evolution
with resonant amplification for wavenumber $k$
between $t_0$ and $t_2$ yields
\begin{equation}
\label{growth}
\frac{\delta X_k(t_2)}{\delta X_{k}(t_0)}\simeq
\left[\frac{\Phi(t_0)}{\Phi(t_2)}\right]^{1/2}
\frac{\exp\left[{2\pi\mu (N_1+N_2)}\right]}{a^{3/2}(t_2)}\, .
\end{equation}

The contribution of $\chi$ fluctuations to the large-scale metric
fluctuations at $t_2$, if estimated via Eqs.~(\ref{explicit})
and (\ref{growth})  and
employing the severely damped background $\chi$, would be utterly 
negligible. Nevertheless, the problem is intrinsically non-linear since
$\delta\chi_k\gg \chi$. If instead, we were to include the second
order contribution such as $\sim(g\varphi)^2\int d^3{\bf k^\prime}
\delta\widetilde{\varphi}_{I{\bf k^\prime}}
\delta\widetilde{\varphi}_{I({\bf k-k^\prime})}$,
on the R.H.S. of Eq.~(\ref{extra}) we would
find significant metric perturbations when backreaction
becomes important, even in the limit $k\to 0$,
sourced by the sub-horizon fluctuations in $\delta\chi$. This is not surprising
though, at onset of backreaction there is approximate equipartition of
energy between the inflaton and the coupled boson. Since the energy in the
$\delta\chi$ fluctuations is not taken into account in Eq.~(\ref{Friedmann}), 
one expands effectively around an unphysical Hubble constant.
We may rather attempt to estimate 
$\phi_{k\to 0}(t_2)$ by associating an effectively
homogeneous background of the $\chi$ field on large scales by taking the
root mean square $X_{rms}$
over small-scale fluctuations. This would yield
$\phi_{k\to 0}|_{\chi}(t_2)\sim g^2\Phi^2 X_{rms}\delta X_k/
g^2\Phi^2 X_{rms}^2 \sim \delta X_k/ X_{rms}\,(t_2)$, for the contribution
of $\chi$ fluctuations to the metric perturbations.
Using Eq. (\ref{amplitude}) with $T_{re}\simeq 1$eV,
$T_0\simeq 4\times 10^{-4}M_{pl}$, and $\varphi_{ex}\simeq 3M_{pl}$
we find $\delta X_k(t_0)\sim 10^{-47}$.
Such a small initial amplitude
is only amplified by modest factors $\sim 10^4-10^5$, estimated via 
Eq. (\ref{growth}) for $\mu\simeq 0.13$ and $g\sim
3\times 10^{-4} - 10^{-2}$.Nevertheless, we will see below that
this result represents a gross underestimate.

Second order contributions to the energy density from scalar field and metric
fluctuations on scales exceeding the typical fluctuation scale
may be accounted for by the effective energy-momentum
tensor formalism of Ref.~\cite{ABM97}. This corresponds to
substituting the potential $\tilde{V}\equiv V+\frac{1}{2}
\sum_{IJ}V_{\varphi_I\varphi_J}\delta\varphi_I\delta\varphi_J$
and adding the term $[\delta\dot{\varphi}_I^2+
a^{-2}(\nabla\delta\varphi_I)^2]/2+
2V_{\varphi_I}\phi\delta\varphi_I$ for each scalar
field under the parentheses in Eq.~(\ref{Friedmann}), and
using the modified equation
\begin{eqnarray}
\label{homogeneous2}
&&\left(\ddot{\varphi}_I + 3H\dot{\varphi}_I\right)
(1+4\phi^2)+\tilde{V}_{\varphi_I}
-2\phi\delta\ddot{\varphi}_I-4\dot\phi\delta\dot{\varphi}_I\nonumber\\
&&-6H\phi\delta\dot{\varphi}_I+4\dot{\varphi}_I\dot{\phi}\phi
-2a^{-2}\phi\nabla^2\delta\varphi_I=0
\end{eqnarray}
for the homogeneous fields.
In these expressions second order terms are to be
understood as spatial averages. 
Whereas for small scales $\sim k_{max}^{-1}$
such a scheme may be questionable, it should be appropriate for the 
super-horizon scales relevant for structure formation. 

\begin{figure}[ht]
\epsfxsize=8.5cm
\hbox to\hsize{\hss\epsfbox{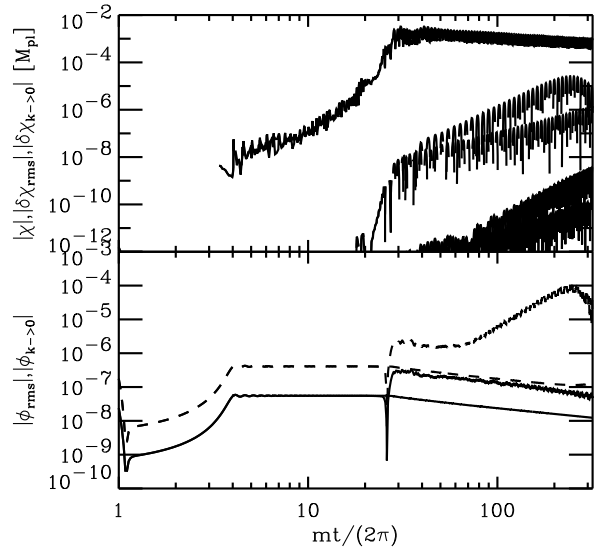}\hss}
\medskip
\caption[...]{\label{F1} Results of a numerical simulation of
metric preheating with two scalar fields and parameters as
discussed in the text. The top panel shows, from top to
bottom, the total $|\delta\chi_{rms}|$, the
homogeneous field $|\chi|$, and the
large scale fluctuations $|\delta\chi_{k\to0}|$. The curve for
$|\delta\chi_{rms}|$ starts at its vacuum expectation value
at the beginning of preheating. The bottom panel
shows the total $|\phi_{rms}|$ (dashed
lines) and the large scale metric potential $|\phi_{k\to0}|$
(solid lines) as functions of time. The results obtained from
integrating Eq.~(\ref{extra}) (upper set of curves) and
Eq.~(\ref{constraint}) (lower set of curves) diverge
once backreaction sets in, reflecting the uncertainties
within the current numerical approach.}
\end{figure}

{\it Numerical Model.} 
We have numerically integrated Eqs.~(\ref{delphi})
and and~(\ref{constraint})
along with Eq.~(\ref{homogeneous2}) for the homogeneous fields
and the Friedman equation~(\ref{Friedmann}) with backreaction terms
for the two field case
and the potential Eq.~(\ref{potential}). We started integration
deep in the inflationary regime where the large scale structure
modes relevant today were still within the Hubble horizon.
We used the initial conditions $\varphi = 4M_{pl}$, $\chi = m/g$ and
Eq.~(\ref{vacampli}) for the matter
fields, as well as $\delta\dot{\varphi}_{Ik}\sim\omega_{Ik}\delta\varphi_{Ik}$
for fluctuation derivatives (the results are qualitatively insensitive
to the exact values). The initial value of
$\phi_k$ was evaluated from
Eq.~(\ref{explicit}). A grid of
100 equidistant logarithmic momentum modes was implemented
between $k=m/2^{1/2}$ and $k=1.5[gm\Phi(t_0)]^{1/2}=3m(q_0/4)^{1/4}$
which covers the dominant resonant modes around the Hubble
horizon. In addition, one mode
with comoving scale of $\sim100\,$Mpc today, corresponding
to $T_{re}\simeq0.5\,$eV in Eq.(\ref{amplitude}), was followed.

In Fig.~\ref{F1} we show results of simulations with
$m=10^{-6}M_{pl}$ and $q_0=10^4$. Note that Fig.~\ref{F1} includes
results from two simulations (a) as outlined above and (b) as above but 
by integrating
Eq.~(\ref{extra}) instead of Eq.~(\ref{constraint}). In the linear limit 
(i.e. $mt/(2\pi)\lesssim25$) the results of both simulations coincide,
providing a consistency check of our numerical routine.
During the inflationary phase one may observe the standard growth of  
metric fluctuations  $\phi$ due to the increase of $(1+w)$, where $w$ is
the ratio of pressure and density. For $mt/(2\pi)\gtrsim5$
parametric resonance develops and for $mt/(2\pi)\gtrsim25$
backreaction becomes important. The saturation of $X_{rms}$ at $\simeq m/g=
q_0^{-1/2}\Phi(t_0)/2\simeq 3\times 10^{-3}$ due to backreaction
is clearly visible in Fig.~\ref{F1}. Note that due to the
choice of $q_0$ and $m$,
$X_{rms}(t_2)\simeq X_{rms}(t_1)$ since $q_1\simeq 1/4$, thus the second
stage of preheating is very short.
When backreaction becomes important, results
obtained from integrating Eq.~(\ref{extra}) and Eq.~(\ref{constraint}),
respectively, which are equivalent in the linear regime,
start to differ. In particular, whereas Eq.~(\ref{extra}) predicts a second
rise in small scale metric fluctuations at $t_2$, and oscillatory behavior,
integration of Eq.~(\ref{constraint}) 
shows a smooth transition. These differences are likely due to the 
replacement of small scale fluctuations by numerical phase space sums
which may introduce some coherence between the perturbed quantities
which in reality, in particular for large scales, does not exist.
Within this approximative scheme, $\phi_{k\to 0}|_{\chi}(t_2)$ is still
$\lesssim \phi_{k\to 0}|_{\varphi}(t_2)$, but much larger than
our \lq\lq naive\rq\rq\ analytical estimate. The coupling
to small-scale metric $\phi$ and boson perturbations $\delta \chi$
in Eq.~(\ref{homogeneous2}) tends to lift the
homogeneous $\chi$ field to $\sim\phi\delta X$ which in turn induces
a source term for $\delta\chi_{k\to 0}$ in Eq.~(\ref{delphi}).
These source terms
induce a growth of $\chi$ and $\delta\chi_{k\to 0}$, with growth due
to parametric resonance subdominant. In contrast, the employed 
model approximates growth of $\delta\chi_{k\simeq k_{max}}$, as well
as backreaction on $\varphi$, by the usual parametric resonance 
theory~\cite{KLS97}, i.e. essentially unmodified by the existence of metric
fluctuations. Our conclusions, 
i.e. $\phi_{k\to 0}|_{\chi}\lesssim \phi_{k\to 0}|_{\varphi}$ and
$\phi_{k\to 0}\ll 1$, are not likely modified by a subsequent epoch
of rescattering
and thermalization, but may be if $\phi$ grows further by gravitational
instability. Nevertheless, for $\phi\sim 1$ efficient formation of primordial
black holes seems inevitable and the notion of an effectively homogeneous
and isotropic universe on large scales, described by a FRW metric,
is called into question.  
   
{\it Summary.}
We have investigated preheating after an inflationary
epoch for a simple chaotic inflationary model with the inflaton
coupled to another bosonic particle.
Our study focuses on the possibility of generation of secondary
metric perturbations on scales relevant for structure formation.
Within first-order perturbation theory, but with backreaction included 
to second order in the scalar fields and in the metric in an effective way,
we found that metric fluctuations on scales relevant for structure
formation stay well within the linear regime during preheating.
This is mainly due to the
fact that at the beginning of preheating the homogeneous component 
of the scalar field $\chi$, and its super-horizon
quantum fluctuations $\delta\chi$, which couple 
to the inflaton, are severely suppressed
compared to the inflaton~\cite{remark}.
Our conclusions are similar to those of 
Refs.~\cite{FB99,PE99} who investigated the case of
general-relativistic parametric resonance of perturbations around
the coherent oscillations of a massive, but otherwise uncoupled inflaton.
However, we caution
that our study does not fully address the possibility of efficient mode-mode
coupling between sub-horizon and super-horizon metric fluctuations.  
This issue may, in principle, be resolved by numerical simulation as in
Ref.~\cite{PE99}. 

{\it Acknowledgments.} We thank Bruce Bassett, Robert Brandenberger, 
David Kaiser, Roy Maartens, and Viatcheslav Mukhanov for helpful
discussions. GS was supported, in part, by the DoE, NSF, and NASA at
the University of Chicago.


\begin{thebibliography}{99}
\bibitem{remark} This was independently realized in a revised
version of Ref.~\cite{BTKM99}. However, regarding the impact
on large scale metric perturbatons, our conclusions differ
from Ref.~\cite{BTKM99}.
\end{thebibliography}
\end{document}